\documentclass[preprintnumbers]{revtex4}
\usepackage{amsmath}
\usepackage{amssymb}
\usepackage{graphicx}

\input{tcilatex}
\begin{document}

\date{\today }
\title{Trion and Biexciton in Monolayer Transition Metal Dichalcogenides}
\author{Roman Ya. Kezerashvili$^{1,2}$, Shalva M.Tsiklauri$^{3}$}
\affiliation{\mbox{$^{1}$New York
City College of Technology, The City University of New York, USA} \\
$^{2}$The Graduate School and University Center, The City University of New
York, USA\\
$^{3}$Borough of Manhattan Community College, The City University of New
York, USA}

\begin{abstract}
We study the trion and biexciton in transition metal dichalcogenides
monolayers within the framework of a nonrelativistic potential model using
the method of hyperspherical harmonics (HH). We solve the three- and
four-body Schr\"{o}dinger equations with the Keldysh potential by expanding
the wave functions of a trion and biexciton in terms of the antisymmetrized
HH. Results of the calculations for the ground state energies are in good
agreement with similar calculations for the Keldysh potential and in
reasonable agreement with experimental measurements of trion and biexciton
binding energies.
\end{abstract}

\maketitle

\section{Introduction}

Monolayer transition metal dichalcogenides (TMDCs) are a new class of
two-dimensional (2D) materials with remarkable optical and electronic
properties. The TMDC family includes MoS$_{2}$, MoSe$_{2}$, WS$_{2}$ and WSe$%
_{2}$, all of which share similar properties with respect to atomic and
electronic structure. Unlike graphene, these 2D crystals are believed to be
direct band gap semiconductors. A result of reduced dimensionality and weak
dielectric screening in such materials is a strong electrostatic interaction
allowing the formation of bound state complexes of electrons and holes with
very large binding energies. The latter phenomenon is remarkably pronounced
in monolayer TMDCs, leading to the formation of tightly bound excitons with
binding energies of several hundred millielectronvolts. An observed
consequence of reduced dimensionality and weak dielectric screening in such
materials is a strong electrostatic interaction allowing the existence of
other stable bound states consisting of a larger number of electrons and
holes, such as positively or negatively charged trions (X$^{\pm }$) and
biexcitons. In TMDC monolayers, X$^{\pm }$ is formed by an exciton with an
extra hole or electron, which can be introduced in different ways. The trion
binding energies extracted from recent\emph{\ }experimental observations
such as photoluminescence, electroluminescence, and absorption spectroscopy
in monolayer TMDCs were found to be in the range of 10-43 meV \cite%
{MoS23Heinz, MoSe21 Ross, MoSe2Singh, ShangBiexiton, WS2Plechinger,
ZhangMS2, WS2ZHU2015, WS2 Zhu, WSe2 Jones, WSe2Wang}. Very recent evidence
of stable bound states of two electrons and two holes---biexcitons---with
binding energies of $\backsim $ 20--70 meV in TMDCs has been reported in
Refs. \cite{ShangBiexiton, WS2Plechinger, MaiBiexiton, MoSe2Hao,
Reichman2015, SieBiexitonMoS2}.

Until now several approaches have been proposed for evaluating the binding
energies of exciton complexes such as trion and biexiton in two-dimensional
transition metal dichalcogenides. Initial work on exciton and trion binding
energies in TMDCs employed variational wave functions \cite{Reichman2013},
and more recently used more intricate trial wave functions \cite%
{Reichman2015, Prada}. Exciton complexes in low dimensional TMDCs studied
using the time-dependent density-matrix functional theory \cite%
{TimeDepdensity matrix functional theory}, the stochastic variational method
using the explicitly correlated Gaussian basis \cite{VargaNano2015,
VargaPRB2016}. Within the effective mass approach, quantum Monte Carlo
methods, such as the diffusion Monte Carlo and the path integral Monte
Carlo, provide accurate and powerful means for studying few-particle
systems. Trions and biexcitons in 2D TMDC sheets of MoS$_{2}$, MoSe$_{2}$, WS%
$_{2}$, and WSe$_{2}$ are studied by means of the density functional theory
and path integral Monte Carlo method in \cite{DenFuncTheoryPIMC}, the path
integral Monte Carlo methodology in \cite{Saxena}, and the diffusion Monte
Carlo approach in \cite{BerkelbachDifMonteCarlo}.

In this work we study the trion and biexciton in TMDC monolayers in the
effective mass approximation within the framework of a nonrelativistic
potential model using the method of hyperspherical harmonics (HH). For the
solution of three- and four-body Schr\"{o}dinger equations with the Keldysh
potential \cite{Keldysh}, we expand the wave functions of three- and four
bound particles in terms of the antisymmetrized hyperspherical harmonics,
and obtain the corresponding hyperradial equations that are solved
numerically.

\section{Theoretical model}

Within the effective mass approximation, the nonrelativisic Hamiltonian of
an excitonic few-particle system lying in a 2D plane is

\begin{equation}
H=-\frac{\hbar ^{2}}{2}\overset{N}{\underset{i=1}{\sum }}\frac{1}{m_{i}}%
\nabla _{i}^{2}+\overset{N}{\underset{i<j}{\sum }}V_{ij}(\left\vert \mathbf{r%
}_{i}-\mathbf{r}_{j}\right\vert ),  \label{hamilton}
\end{equation}%
where $m_{i}$ and $\mathbf{r}_{i}$ are the effective mass and the $i$th
particle position, respectively. We assume only two types of charge
carriers: electrons and holes with the corresponding effective masses. Below
we restrict ourselves to $N=3$ (trion) and $N=4$ (biexciton). The screened
Coulomb interaction $V_{ij}(\left\vert \mathbf{r}_{i}-\mathbf{r}%
_{j}\right\vert )$ between $q_{i\text{ }}$ and $q_{j}$ charges in Eq. ~(\ref%
{hamilton}) for monolayer TMDCs was derived by Keldysh \cite{Keldysh}:

\begin{equation}
V_{ij}(r)=\frac{\pi q_{i}q_{j}}{\rho _{0}}\left[ H_{0}(\frac{r}{\rho _{0}}%
)-Y_{0}(\frac{r}{\rho _{0}})\right] .  \label{poten}
\end{equation}%
\ \ \ In Eq.~(\ref{poten}) $H_{0}(\frac{r}{\rho _{0}})$ and $Y_{0}(\frac{r}{%
\rho _{0}})$ are the Struve function and Bessel function of the second kind,
respectively, $\rho _{0\text{ }}$ is the screening length $\rho _{0}=2\pi
\chi ,$ where $\chi $ is the polarizibility of the $2$D materials, which
sets the boundary between two different behaviors of the potential due to a
nonlocal macroscoping screening. For large distances $\ r>>\rho _{0\text{ }}$%
the potential has the three-dimensional Coulomb tail, while at very \ small $%
r<<\rho _{0\text{ }}$distances it becomes a logarithmic Coulomb potential of
a point charge in two dimensions. \ A crossover between these two regimes
takes place around distance $\rho _{0}$.

To obtain a solution of the Schr\"{o}dinger equations for the trion and
biexciton using the Hamiltonian (\ref{hamilton}), we use the method of
hyperspherical harmonics. The main idea of this method is the expansion of
the wave function of the corresponding excitonic states in terms of \ HH
that are the eigenfunctions of the angular part of the Laplace operator in
the four-dimensional (4D) space (trion) or in the six-dimensional (6D) space
(biexciton). As the first step by introduction of the trees of Jacobi
coordinates for a trion or biexciton and considering that the electron and
hole have unequal masses one can separate the center-of-mass and write the
nonrelativistic Schr\"{o}dinger equation for the relative motion of $N$
particles. The next step is the introduction of the hyperspherical
coordinates in the 4D space for the trion or in the 6D space for the
biexciton and one introduces the hyperspherical coordinates in 2$(N-1)$%
-dimensional configuration space, given by the hyperradius $\rho ^{2}$ = $%
\sum\limits_{l=1}^{N}x_{l}^{2},$ where $x_{l}$ are Jacobi coordinates, and a
set of angles $\Omega _{\rho }$ \cite{Avery, JibutiSh}, which define the
direction of the vector $\mathbf{\rho }$ in $2(N-1)$-dimensional space and
rewrite in hyperspherical coordinates the Schr\"{o}dinger equation for the
relative motion of $N-$particles. By expanding the wave function of $N$
bound particles in terms of the HH one obtains 
\begin{equation}
\Psi (\rho ,\Omega _{\rho })=\rho ^{-\frac{2N-3}{2}}\underset{\mu \text{ }%
\lambda }{\sum }u_{\mu }^{\lambda }(\rho )\Phi _{\mu }^{\lambda }(\Omega
_{\rho },\mathbf{\sigma }),  \label{HH1}
\end{equation}%
where $\Phi _{\mu }^{\lambda }(\Omega _{\rho },\mathbf{\sigma })$ are fully
antisymmetrized functions with respect to two electrons in the case of the
negative trion and two electrons and two holes in the case of the biexciton.
These functions are constructed from spin function and the hyperspherical
harmonics. The HH are the eigenfunctions of the angular part of the $2(N-1)$%
-dimensional Laplace operator in configuration space with eigenvalue $%
L_{N}(L_{N}+1)$, where $L_{N}$ $=\mu +(2N-5)/2$. They are expressible in
terms of spherical harmonics and Jacobi polynomials \cite{Avery, JibutiSh}$.$
In Eq. (\ref{HH1}), for the sake of simplicity, we denote by $\lambda $ the
totality of quantum numbers on which the $N-$body hyperspherical harmonics
depend and the integer $\mu $ is the global momentum in the $2(N-1)$%
-dimensional configuration space, which is the analog of angular momentum in
the case of the exciton, $N=2$. \ Introducing the expansion (\ref{HH1}) in
the Schr\"{o}dinger equation for $N-$bounded particles one can separate the
radial and angular variables that results in a system of coupled
differential equations for the hyperradial functions\ $u_{\mu }^{\lambda
}(\rho )$

\begin{equation}
\frac{d^{2}u_{\mu }^{\lambda }(\rho )}{d\rho ^{2}}+\left[ \kappa ^{2}-\frac{%
L_{N}(L_{N}+1)}{\rho ^{2}}\right] u_{\mu }^{\lambda }(\rho )=\underset{\mu
^{^{\prime }}\lambda ^{^{\prime }}}{\sum }V_{\mu \mu ^{^{\prime }}\text{ }%
\lambda \lambda ^{^{\prime }}}(\rho )u_{\mu ^{^{\prime }}}^{\lambda
^{^{\prime }}}(\rho ),  \label{HH2}
\end{equation}%
where%
\begin{equation}
V_{\mu \mu ^{^{\prime }}\text{ }\lambda \lambda ^{^{\prime }}}(\rho )=\frac{%
2M}{\hslash ^{2}}\int \text{ }\left[ \Phi _{\mu }^{\lambda }(\Omega _{\rho },%
\mathbf{\sigma })\right] ^{\ast }\left( \sum\limits_{i<j}V_{ij}\right) \Phi
_{\mu ^{^{\prime }}}^{\lambda ^{^{\prime }}}(\Omega _{\rho },\mathbf{\sigma }%
)d\Omega _{\rho }
\end{equation}%
is the $N-$particle effective potential energy defined by the Keldysh
potential $V_{ij}$ (\ref{poten}), $\kappa ^{2}=2MB/\hslash ^{2}$, where $B$
is the binding energy, and $M$ is a reduced mass for trion or biexciton$.$

\section{Results of calculations}

The system of coupled differential equations (\ref{HH2}) for the hyperradial
functions\ $u_{\mu }^{\lambda }(\rho )$ is infinite and the corresponding
hyperradial equations are solved numerically. By solving the system of
equations (\ref{HH2}) one finds the binding energy as well as the
corresponding hyperradial functions. The latter allows one to construct the
wave function $\Psi (\rho ,\Omega _{\rho })$ (\ref{HH1}). Reasonable
convergence is reached for $\mu _{max}$ = 10 and we limit our considerations
to this value. In calculations we use the necessary parameters for the trion
and biexciton Hamiltonians that were calculated from first principles. The
resulting binding energy of excitonic systems is a function of only $\rho
_{0}$ and the electron-hole mass ratio $m_{e}/m_{h}$. In our calculations we
use the effective masses extracted from the low energy band structure
obtained in the density functional theory \cite{DFTdata} or the GW
approximation \cite{GWdata}, while the screening length $\rho _{0}$ was
calculated using the polarizibility $\chi $ for TMDCs given in Ref. \cite%
{Reichman2013}. The results of our calculations for the binding energy of
the trion and biexciton in MoS$_{2}$, MoSe$_{2}$, WS$_{2}$, and WSe$_{2}$
along with experimental data are presented in Table 1 and Table 2. For
comparison we presented the results of other theoretical studies where the
Keldysh potential \cite{Keldysh} was used to find the binding energies of
trion and biexciton. Our TMDC binding energies for the trion and biexciton
agree well with those calculated via the stochastic variational method using
a correlated Gaussian basis \cite{VargaNano2015, VargaPRB2016}, the path
integral Monte Carlo \cite{Saxena}, diffusion Monte Carlo \cite%
{BerkelbachDifMonteCarlo} and density functional theory and path integral
Monte Carlo \cite{DenFuncTheoryPIMC} methods. In average, the discrepancies
are less than $\pm $1 meV. However, there is significant disagreement with
the variational calculations \cite{Reichman2013}. There is a discrepancy
with experiment for the biexciton case for MoS$_{2}$, WS$_{2}$, and WSe$_{2}$
with all theoretical predictions, while the recent experimental result for
MoSe$_{2}$ \cite{MoSe2Hao} is in reasonable agreement with our calculation
and theoretical results \cite{VargaNano2015, VargaPRB2016,
DenFuncTheoryPIMC, Saxena, BerkelbachDifMonteCarlo}.

\begin{table}[t]
\caption{Experimental and theoretical results for negative trion binding
energies in meV for TMDCs materials. The abbreviations are the following: V-
Variational Method; SVM - Stochastic Variational Method; PIMC - Path
Integral Monte Carlo Method; DFT \& PIMC - Density Functional Theory and
Path Integral Monte Carlo Method; DMC - Diffusion Monte Carlo Method.}
\label{tab1}
\begin{center}
\begin{tabular}{cccccccc}
\hline\hline
TMDC & Present work & Experiment & V \cite{Reichman2013} & SVM \cite%
{VargaNano2015, VargaPRB2016} & PIMC \cite{Saxena} & DFT \& PIMC \cite%
{DenFuncTheoryPIMC} & DMC \cite{BerkelbachDifMonteCarlo} \\ \hline
MoS$_{2}$ & 32.8 & 18$\pm $1.5 \cite{MoS23Heinz}, 43 \cite{ZhangMS2} & 26 & 
33.7 &  & 32.0 & 33.8 \\ 
MoSe$_{2}$ & 27.6 & 30 \cite{MoSe21 Ross, MoSe2Singh} & 21 & 28.2 &  & 27.7
& 28.4 \\ 
WS$_{2}$ & 33.1 & 10-15 \cite{ShangBiexiton}, 30 \cite{WS2Plechinger}, 34 
\cite{WS2ZHU2015}, 45 \cite{WS2 Zhu} & 26 & 33.8 & 28 & 33.1 & 34.0 \\ 
WSe$_{2}$ & 28.3 & 30 \cite{WSe2 Jones, WSe2Wang} & 22 & 29.5 &  & 28.5 & 
29.5 \\ \hline\hline
\end{tabular}%
\end{center}
\end{table}

\begin{table}[t]
\caption{Experimental and theoretical results for biexciton binding energies
in meV for TMDCs materials. Notations are the same as in Table 1.}
\label{tab2}
\begin{center}
\begin{tabular}{ccccccc}
\hline\hline
TMDC & Present work & Experiment & SVM \cite{VargaNano2015, VargaPRB2016} & 
PIMC \cite{Saxena} & DFT \& PIMC \cite{DenFuncTheoryPIMC} & DMC \cite%
{BerkelbachDifMonteCarlo} \\ \hline
MoS$_{2}$ & 22.1 & 40, 60 \cite{SieBiexitonMoS2}, 70 \cite{MaiBiexiton} & 
22.5 &  & 22.7 & 22.7 \\ 
MoSe$_{2}$ & 17.9 & $\thicksim $20 \cite{MoSe2Hao} & 18.4 &  & 19.3 & 17.7
\\ 
WS$_{2}$ & 23.1 & 45 \cite{ShangBiexiton}, 65 \cite{WS2Plechinger} & 23.6 & 
21 & 23.9 & 23.3 \\ 
WSe$_{2}$ & 19.8 & 52 \cite{Reichman2015} & 20.2 &  & 20.7 & 20.0 \\ 
\hline\hline
\end{tabular}%
\end{center}
\end{table}

\section{Conclusion}

We have applied the hyperspherical harmonics method to the calculation of
binding energies for three- to four-body excitonic formations in TMDCs. Our
results lie in good agreement with similar theoretical effective mass model
findings for the trions and biexcitons in MoS$_{2}$, MoSe$_{2}$, WS$_{2}$,
and WSe$_{2}.$ There is reasonable agreement with the existing experimental
binding energies for the cases of the trion in MoSe$_{2}$, WS$_{2}$, and WSe$%
_{2}$ and the biexciton in MoSe$_{2}$. Our disagreement with the variational
calculations in the case of the trion may be due to its constraint on the
symmetry of the trial wave function. However, our findings for the ground
state energies for the trion and biexciton confirm and agree well with
previous calculations within of the aforementioned approaches where the
Keldysh potential was used. The comparison of our results with existing
calculations performed within different methods allows one to estimate the
accuracy of the methods, and understand the importance of the screened
electron-hole interaction in formation of electron-hole complexes.

\acknowledgments

Sh. M. T is supported by PSC CUNY Grant: award \ No. 69536-00 47. R. Ya. K.
is supported by the NSF Grant Supplement to the NSF Grant No. HRD-1345219.

\end{document}